\documentclass{appolb}
\usepackage{graphicx}
\usepackage{subfigure}
\usepackage{eqnarray}
\usepackage{color}

\begin{document}
\title{TRI$\mu$P - A new facility to produce and trap radioactive isotopes%
\thanks{Presented at the XXIX Mazurian Lakes Conference on Physics, Piaski, Poland,
August 30 - Septenber 6, 2005}%
}
\author{M. Sohani
\address{}
\normalsize{for the TRI$\mu$P group \cite{TRImP}}
\address{Kernfysisch Vensneller Instituut, Zernikelaan 25, 9747 AA Groningen, the Netherlands}
}
\maketitle
\begin{abstract}
At the Kernfysisch Vensneller Institiutr (KVI) in Groningen, NL, a
new facility (TRI$\mu$P) is under development. It aims for
producing, slowing down, and trapping of radioactive isotopes in
order to perform accurate measurements on fundamental symmetries
and interactions. A production target station and a dual magnetic
separator installed and commissioned. We will slow down the
isotopes of interest using an ion catcher and in a further stage a
radiofrequency quadropole gas cooler (RFQ). The isotopes will
finally be trapped in an atomic trap for precision studies.
\end{abstract}

\section{Introduction}
Rare and short lived radioactive isotopes are of interest because
they can offer unique possibilities for investigating fundamental
physical symmetries \cite{HA04}.
 Fundamental symmetries are at the basis of the Standard Model (SM).
 Using radioactive isotopes, limits for the validity of the SM can be explored in high precision measurements.
 In particular, high accuracy can be achieved, when suitable radioactive isotopes are stored in atom or ion traps
 \cite{WI03,JU05}.

The TRI$\mu$P (Trapped Radioactive Isotopes: $\mu$icrolaboratories for fundamental Physics)
 facility at the Kernfysisch Vensneller Instituut(KVI) in Groningen, The Netherlands,
 is being developed to conduct such high precision studies. The
 local group concentrates on precision measurements of nuclear
 $\beta$-decays \cite{JU05-2} and search for permanent electric
 dipole moments \cite{JU05-2}.
 We will briefly describe the complete facility consisting of
 a production target, a Magnetic Separator with cooling stages and atom traps. We will describe also the method of
 the precision measurements in beta-decay studies.

\section{Magnetic Separator}
\begin{figure}
\centering
\includegraphics[width=\textwidth]{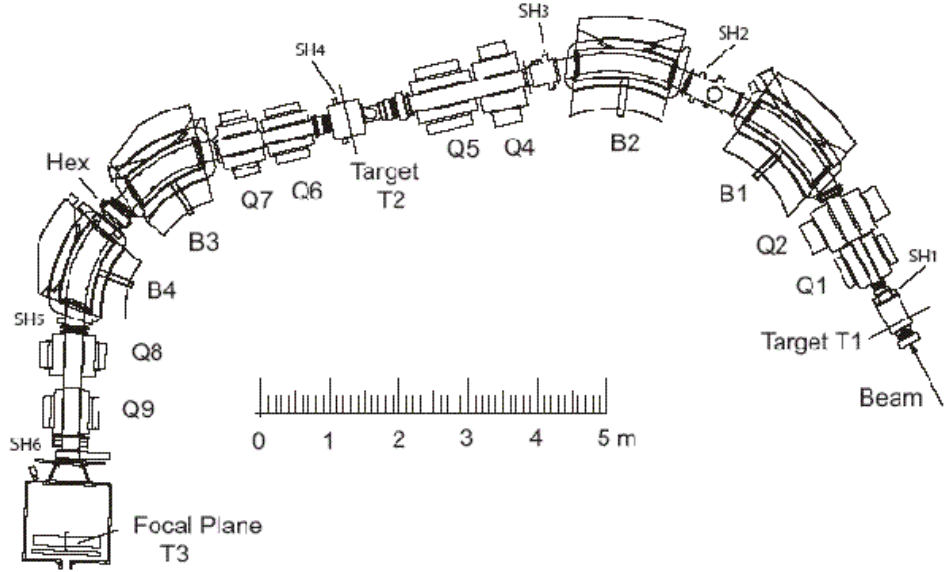}

\caption{Layout of the TRI$\mu$P separator.}
\label{fig:layout}
\end{figure}

Heavy-ion beams from the superconducting cyclotron AGOR at KVI are
used to produce a wide range of products(Fig.~\ref{fig:layout}).
 A hydrogen gas target cooled to liquid nitrogen temperature \cite{YO05} and various solid
targets have been employed in different types of reactions,
 from fusion evaporation to charge exchange. Using inverse kinematics, products are
 selected by the dual magnetic separator \cite{BE05}.

\begin{figure}[htbp]
\centering \begin{minipage}[c]{0.5\textwidth} \centering
\subfigure[]{ \label{fig:na.eps}
\includegraphics[width=2.75in]{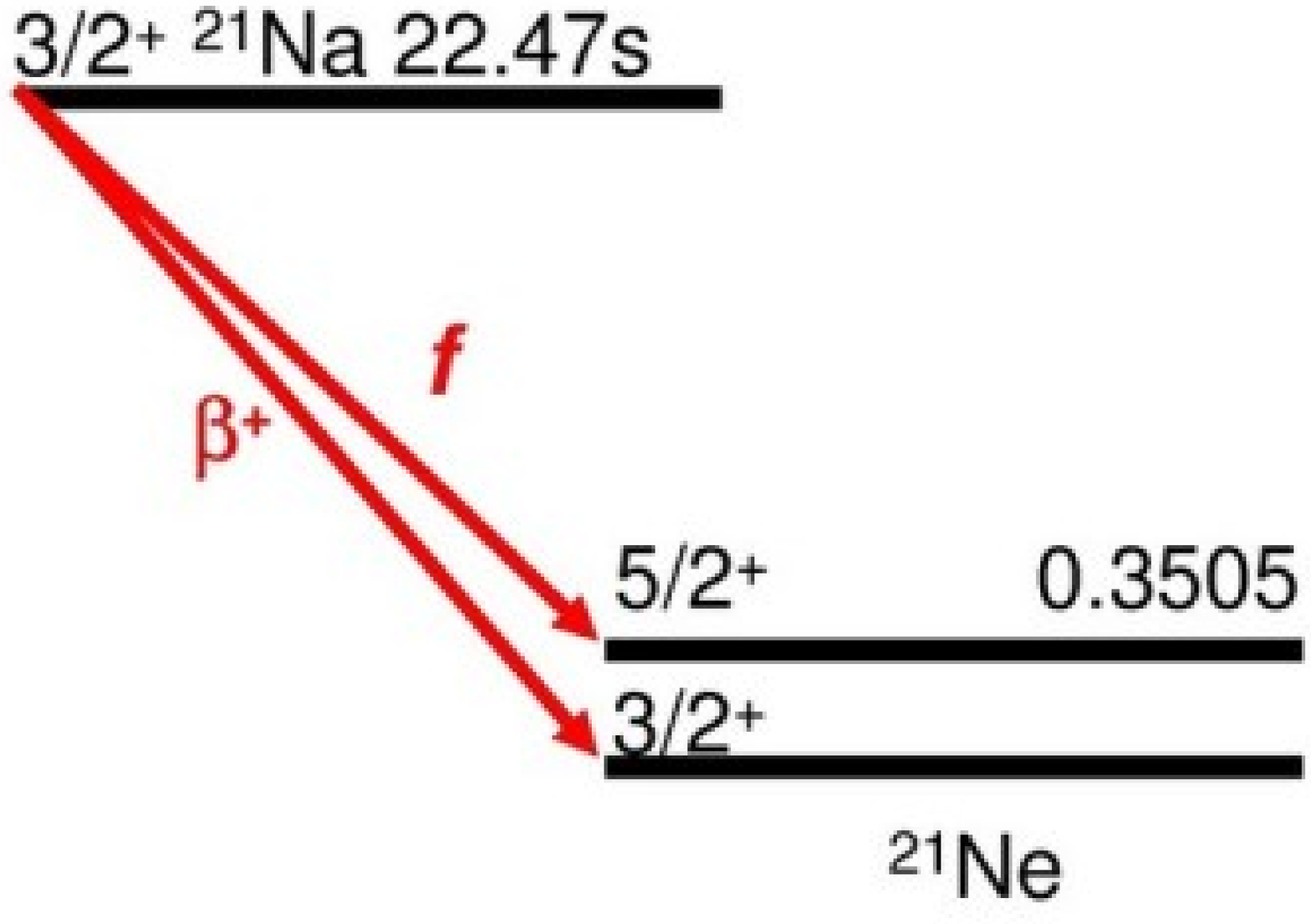}}
\end{minipage}%
\begin{minipage}[c]{0.5\textwidth} \centering \subfigure[]{ \label{fig:LPC}
\includegraphics[width=2.75in]{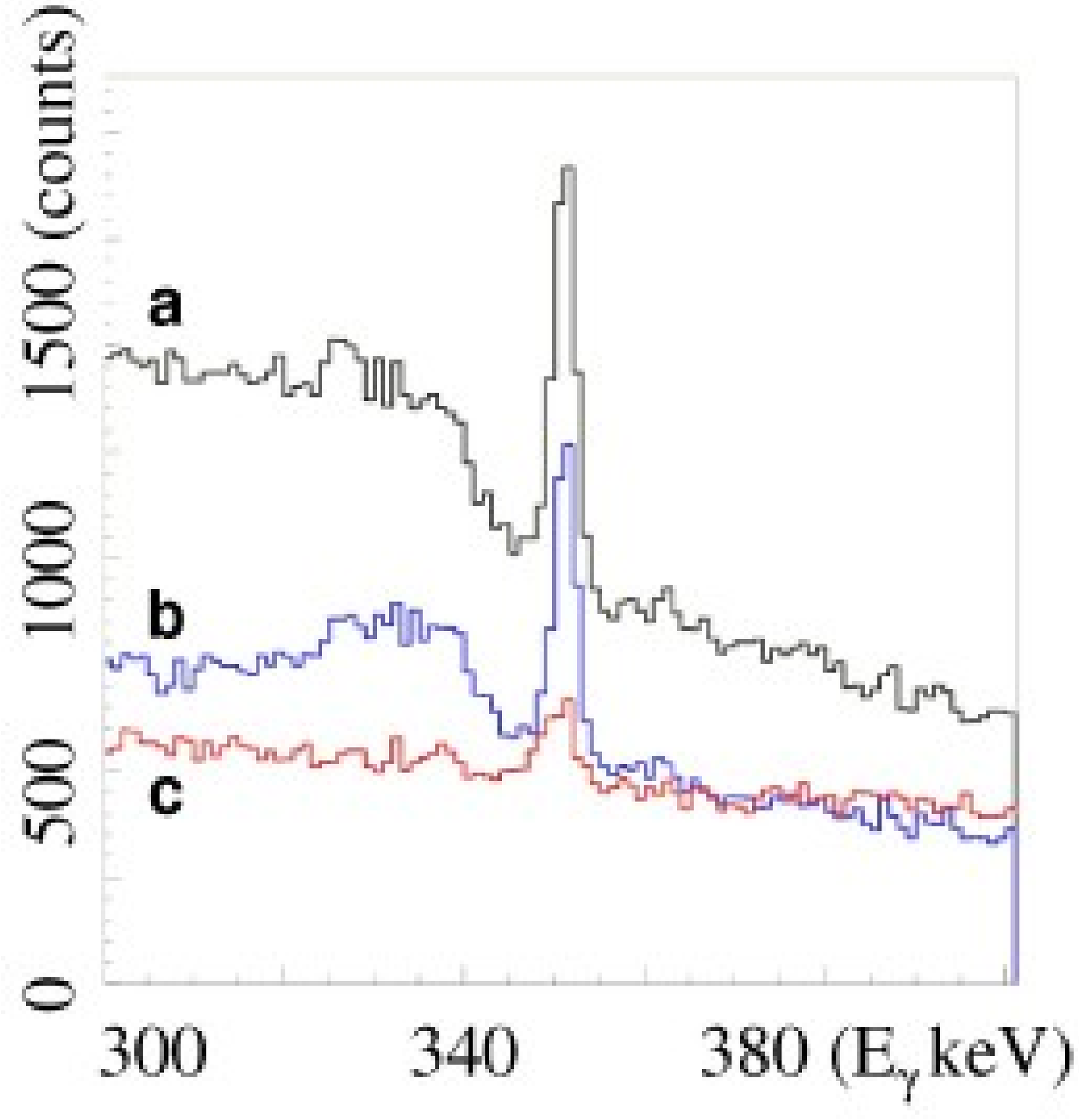}}
\end{minipage}
\caption{a)$^{21}Na$ $\beta$-decay scheme. The experiment aims to
determine the fraction \textit{f} ending in the $\frac{5}{2}^+$
state. b) Observed $\gamma$-ray spectra: a)-total spectrum in
coincident with $\beta$-decay; b)-Compton suppressed spectrum of
(a); c)-Compton suppressed background spectrum}
\end{figure}

The TRI$\mu$P separator has been commissioned.The experiments
included $^{21}$Na production using $^{21}$Ne (43 MeV/n and 20
MeV/n). Typical rates were $3\times10^3$/s/pnA for $^{21}$Na.
Recently $10^4$ /s/pnA for $^{19}$Ne and $10^3$/s/pnA $^{20}$Na
was achieved. In a first physics experiment $^{21}$Na and
$^{22}$Mg were used to measure branching ratio of $^{21}$Na
$\beta$ decay to the exited state and of $^{21}$Ne at 350 keV
\cite{Achouri}. The population of this state is of relevance for
$\beta-\nu$ correlation measurements \cite{SC04}. A stack of two
silicon detectors registered the incoming particles and the
subsequent $\beta$ decay. A set of Ge detectors detected the
$\gamma$ ray emission following part of the $\beta$ decays.
  The Ge Clover detectors include a BGO Compton-shield to reduce background.
 The systematic variation in the branching ratio, that has been observed in various experiments \cite{AL74-WI80},
 may is been caused by a line in the ambient background from the $^{238}$U decay chain (352 keV, $^{214}$Pb).
 This line is clearly seen in the spectrum (see Fig.~\ref{fig:LPC}).

\section{Cooling stages}
In order to perform precision measurements, radioactive ions
produced and separated from the primary beam and unwanted products
need to be cooled and trapped in atom traps. The cooling procedure
needs to be fast and efficient in view of the short lifetime of
isotopes. For light isotopes such as $^{21}Na$, the high energy
and fully stripped secondary beam passes trough several
 cooling stages before it can be delivered as a low energy and singly charged ion beam
 with acceptable emittance. Neutralization and laser trapping is the last step to collect
 radioactive atoms in a well defined cold cloud atoms.

\subsection{Ion Catcher}
 The main principal of an Ion Catcher is to stop high-energy ions in matter, i.e. in a gas or a solid.
 Electronic stopping causes fast slowing. In order to extract a low energy ion beam at the end of the ion catching process,
 the particles must remain ionized. Neutralization of atoms is a
 poisoning process which must be avoided. The commonly used techniques
   include gas-filled ion catchers and thermal ionizing devices.

 \textit{Gas-filled ion catcher:} Collisions and charge exchange processes bring ions
 to lower energies and charge states as long as they move trough the gas.
 At low energies, re-ionization and neutralization are in competition.
 Cross sections of these processes are highly dependent on the energy of the ions and
 the relative ionization potential of the particle and the stoping
 gas. There are two strategies: one either relies on the survival
 of the particles as singly charged ions or on active re-ionization
 after neutralization.
 In the latter case extraction can be done e.g. after resonant laser ionization
 \cite{HU02}.
 Principally, the gas-filled ion catcher can only work
 when the space charge built up during stopping does not hinder the extraction optics.
 Therefore, the efficiency for this device depends on the input ion rate \cite{HU02,MOpr}.

 \textit{Thermal ionizer:}  a device consists of one or several metal foils that stop the beam.
 Atoms will defuse out of the foils particularly at high temperature.
 Collisions with the surface of the foils and the cavity body can ionize these atoms.
 A potential electric field allows to extract ions.
 Choosing proper material such as W for the stopping foils and the cavity is important.
 The difference between the ionization potential of the beam element and the foil element determines
 the ionization probability and re-neutralization after each collision with the surface.
 Therefore a thermal ionizer mainly works for alkali and alkali-earth elements that have low ionization potential \cite{KI90}.

 Investigations for both types of devices were performed in the frame work of the TRI$\mu$P program.
 Because of the priority of the experiments on Na and Ra isotopes,
 a thermal ionizer is under the construction and will be tested soon.
 A stack of W foils at high temperature has been chosen as stopper inside a hot W cavity.

\subsection{RFQ Cooler and Buncher}
A segmented radio frequency quadrupole cooler and buncher is the
second stage of the cooling procedure in the TRI$\mu$P facility.
  Input ions coming from the thermal ionizer have an energy spread of order several eV. In the first part of
 the RFQ, they pass through a low-pressure gas-filled medium. A longitudinal drag voltage guides them to the exit
  while confined
 at the same time in the transverse direction by a set of radio frequency electric quadrupoles \cite{LU99}.
 The second part of the RFQ collects ions in a Paul-trap.
 By switching the last electrode of the trap, a bunch of ions can be released and sent to the atom traps through
 an electrostatic low-energy beam line.

The TRI$\mu$P RFQ has been constructed and is commissioned.
Measuring the transmission of the ions and optimizing the settings
are the main issues of the commissioning.

\subsection{Atomic Trap}
The $\beta$-decay experiments st the TRI$\mu$P facility will be
carried out using a set of two subsequent magneto optical traps.
The first trap is to collect atoms and bunch them toward the
second trap inside a detection chamber. The first trap is made of
a small glass cavity and has wide laser beams for efficient
collection of atoms. It will contain a hot Yttrium foil to
neutralize the incoming ion beam. The second trap is located in a
precision measurement chamber centering a reaction microscope and
$\beta$-electron counter.

\section{Beta decay spectroscopy}
Correlations between particles from $\beta$-decay manifest the
symmetries and symmetry violations of the weak interaction
\cite{WI05,SE05}. In weak interactions (a current-current
interaction) several currents can contribute. these are Scalar(S),
Vector (V), Axial-vector (A) and Tensor (T) currents. In the
Standard Model (SM), the weak interaction is exclusively the
result of V and A currents. The V current is observed in Fermi (F)
decay and the A current in Gamow-Teller (GT) decay. Contribution
of other currents in $\beta$-decay will affect the correlations
between the particles and their kinematics. Deviations from the
standard V-A model indicates physics beyond the SM.
\begin{figure}[h]
\centering
\includegraphics[bb=0 0 360 200, scale=0.7]{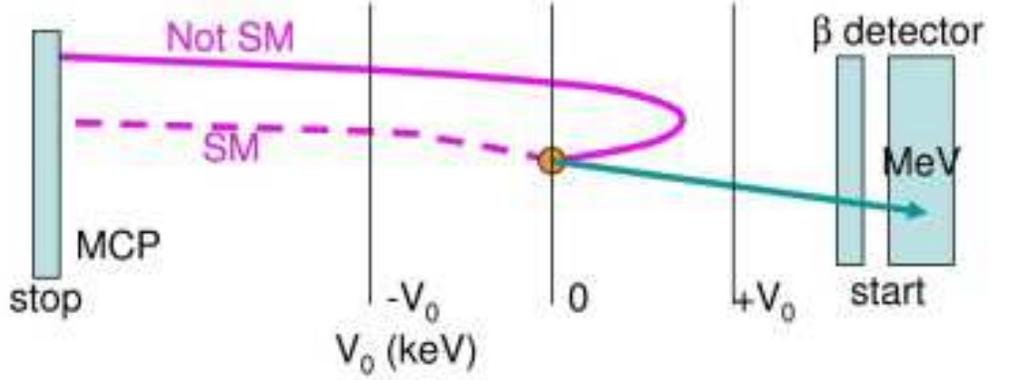}
\caption{RIMS sketch} \label{fig:RIMS}
\end{figure}
It is not practical to measure the $\nu$ particle in $\beta-\nu$
correlations. Therefore, one measures the correlation between the
$\beta$-electron  and recoil ion in order to measure the complete
phase-space of the $\beta$-decay. Recoil-Ion-Momentum Spectroscopy
(RIMS) is an advanced method to measure the recoil ion. This
method is used at KVI to study charge exchange process with the
recoil energy in order of eV \cite{TU01,KN05}. In RIMS the recoil
ions are projected with an electric field on a position sensitive
microchanel plate detector (MCP)(Fig.~\ref{fig:RIMS}). This method
has been used successfully by other groups for $\beta$-decay
studies \cite{SC04,GO00}.

 The time of flight measurement of the recoil ion is started by observing a hit in the $\beta$-detector
 and is stopped by the hit on the MCP.
 Together with the position of the recoil ion on the MCP, this provides information about the initial momentum.
  A position sensitive $\beta$-detector allows to restrict the momentum of the $\beta$-electron.

The differential rate of $\beta$-decay is proportional to the
3-body phase-space and the week interaction matrix element.
 The general derivation of the matrix element (eq.~\ref{equ1}) shows all the possible correlations between the particles \cite{JA57}.

\begin{eqnarray}
  \frac{{\rm d}^2W}{{\color{green} {\rm d}\Omega_e}{\color{red}{\rm d}\Omega_\nu}}
   \sim \!\!& 1 &\!\!\! +\
 a\,\frac{\mbox{\color{green} \boldmath $p$}\cdot{\color{red}\hat{\mbox{\boldmath
$q$}}}}{\color{green} E}
   + b\,\Gamma\,\frac{\color{green} m_e}{\color{green} E} \nonumber \\ \!\!& + &\!\!\!
     \langle\mbox{\boldmath\color{blue} $J$}\rangle\cdot
     \left[ A\,\frac{\mbox{\color{green}\boldmath $p$}}{\color{green} E} +
  B\,{\color{red}\hat{\mbox{\boldmath $q$}}} +
  D\,\frac{\mbox{\color{green} \boldmath $p$}\times
     {\color{red} \hat{\mbox{\boldmath $q$}}}}{\color{green} E}
     \right]  \nonumber \\ \!\!& + &\!\!\!
     \langle\mbox{\color{green} \boldmath $\sigma$}\rangle\cdot
     \left[G\,\frac{\mbox{\color{green}\boldmath $p$}}{\color{green} E} +
     Q\,\langle\mbox{\boldmath\color{blue} $J$}\rangle +
     R\,\langle\mbox{\boldmath\color{blue} $J$}\rangle
     \times\frac{\mbox{\color{green} \boldmath $p$}}{\color{green} E}\right] \
\label{equ1}
\end{eqnarray}

where \textit{p} and \textit{q} are momentum of the
$\beta$-electron and the neutrino, $\langle \it{J}\rangle$ and
$\langle \sigma \rangle$ are polarization of the parent nuclei and
the $\beta$-electron and \textit{E} is the energy of the
$\beta$-electron. \textit{a,b,A,B,D,G,Q and R} coefficients are
depend on the fundamental weak coupling constants and nuclear
matrix elements. The D and R coefficients are zero if time
reversal symmetry is conserved. Experimental searching for finite
value of the D and R coefficients requires a sample of polarized
nuclei ($\langle \it{J}\rangle\neq0$).

For the R coefficient also measuring the polarization of the
 electrons is required. We aim primarily at measuring D.
 In the initial studies without polarization, measuring \textit{a} and \textit{b} coefficient
 explores the effect of the non-SM weak currents.
\begin{figure}[h] \centering
\subfigure[]{
\includegraphics[width=2.75in]{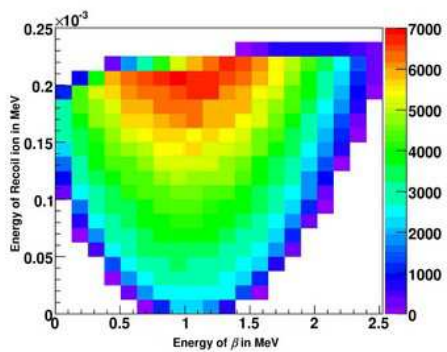}}
\centering \subfigure[]{
\includegraphics[width=2.75in]{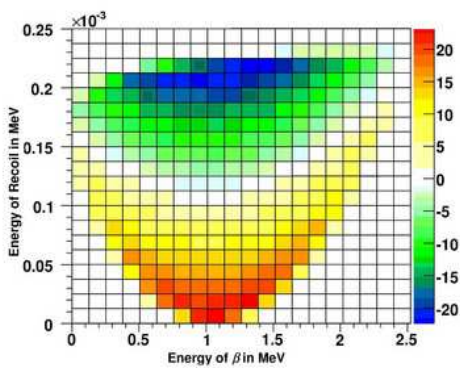}}
\caption{An event distribution according to a MC simulation of
energy of the recoil ion vs energy of the $\beta$-particle. a)
distribution for $^{21}$Na $\beta$-decay. b) difference in the
distribution for a 1$\%$ smaller value of \textit{a} as compared
to the SM prediction. The sensitivity to \textit{a} is largest for
$\beta$ energies around 1 MeV and recoil ion energies around 0.2
MeV.} \label{fig:HIS}
\end{figure}

 In general, there are two free parameters out of 9 parameters for the 3-body phase-space,
 including all the conservation laws and geometrical symmetries for non-polarized atoms.
 Therefore one can choose any pair of the parameters to explore the $\beta$-decay events with a 2-D histogram.
One choice uses the energy of the recoil ion and the
$\beta$-particle as free parameters to express all the other
 parameters in terms of these two.
  A deviation of \textit{a} from the SM prediction results in a change in the corresponding
  distribution as showed in Fig.~\ref{fig:HIS}.

\section{Outlook and conclusion}
  The TRI$\mu$P facility is a user facility to deliver radioactive beams  and to perform precision measurements.
  The TRI$\mu$P production target and the separator are functioning.
  A first user experiment has been already completed.
  Several experiments using a clean radioactive beam have been requested.

\end{document}